# Monitoring the low doping regime in graphene using Raman 2D peak-splits: Comparison of gated Raman and transport measurements

*Zhuofa Chen†, Nathan Ullberg‡, Mounika Vutukuru†, David Barton† and Anna K Swan\*†‡+*

†Department of Electrical and Computer Engineering, Boston University, 8 St Mary's St, Boston Massachusetts 02215, United States of America.

‡Department of Physics, Boston University, 590 Commonwealth Ave, Boston Massachusetts 02215, United States of America.

+Photonics Center, Boston University, 8 St Mary's St, Boston Massachusetts 02215, United States of America.

# TOC

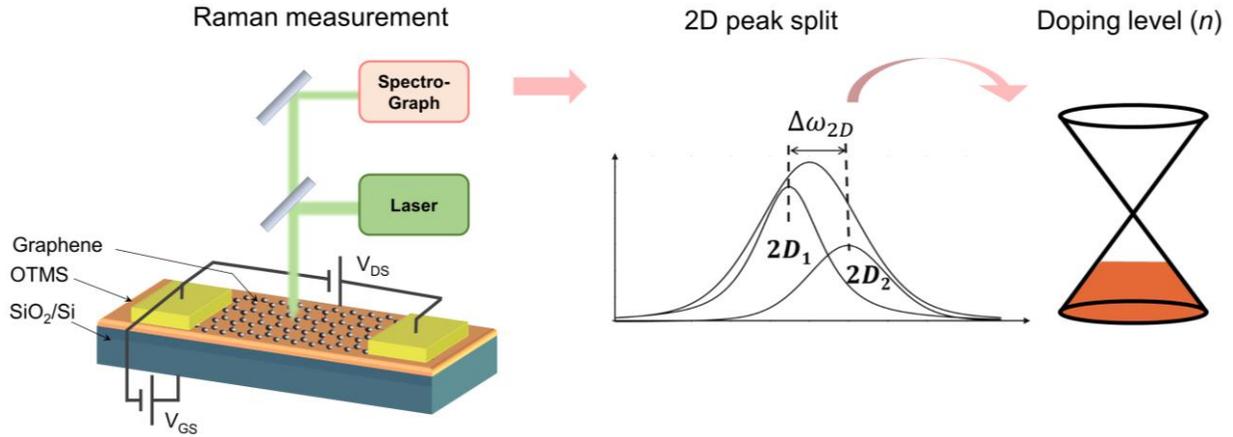

ABSTRACT: Avoiding charge density fluctuations and impurities in graphene is vital for high-quality graphene-based devices. Traditional characterization methods require device fabrication and electrical transport measurements, which are labor-intensive and time-consuming. Existing optical methods using Raman spectroscopy only work for doping levels higher than ~$10^{12}$ cm$^{-2}$. Here, we propose an optical method using Raman 2D peak-splitting (split between the Raman $2D_1$ and $2D_2$ peaks at low doping levels). Electrostatically gated Raman measurements combined with transport measurements were used to correlate the 2D peak-split with the charge density on graphene with high precision (2 ×$10^{10}$ cm$^{-2}$ per 2D peak-split wavenumber). We found that the Raman 2D peak-split has a strong correlation with the charge density at low doping levels, and that a lower charge density results in a larger 2D peak-split. Our work provides a simple and non-invasive optical method to quantify the doping level of graphene from $10^{10}$ cm$^{-2}$ to $10^{12}$ cm$^{-2}$, two orders of magnitude higher precision than previously reported optical methods. This method provides a platform for estimating the doping level and quality of graphene before fabricating graphene devices.

Since its realization in 2004, graphene has attracted a lot of attention due to its superior transport properties such as its giant intrinsic mobility and distinctive electronic structure.[1, 2] The gate-tunable electrostatic charge doping, which is possible for 2D materials, allows for doping without introducing scattering from dopant ions. For the case of graphene, the absence of scattering sites makes for extraordinarily high mobility, which renders graphene an intriguing material for next-generation nanoelectronic devices,[3, 4] including the recently reported unconventional superconductivity in twisted-bilayer graphene that has introduced twistronics as a new platform for applications of graphene.[5] For all high-quality graphene-based devices, reducing charge impurities and charge fluctuations is essential. Therefore, the identification of charge density fluctuations and impurities in graphene has become vital for studying graphene-based applications.

The benchmark method to determine the quality of a device is the electrical transport measurement where defect and charge scattering can be determined. The transport plot (Resistivity vs. Gate voltage) can be analyzed to determine the mobility, the intrinsic charge doping level and the overall charge fluctuations with high precision.[6-8] However, transport measurements require a labor-intensive fabrication process. The objective of our research is to develop an effective optical method to monitor the doping level and charge fluctuations in graphene before any fabrication process.

Graphene deposited directly on $SiO_2$ has high accidental doping levels ($> 10^{12}$ $cm^{-2}$)[9] and charge puddles formed by charged surface states and impurities.[10, 11] Raman spectroscopy has been employed as a noninvasive and efficient method to monitor these high doping levels (accidental charge density) of graphene on substrates.[12-18] Pisana *et al.* studied the Raman G peak of doped graphene and found the correlation between the charge density and the G peak frequency of graphene due to the Kohn anomaly[16] which shifts the G peak up with increasing doping level above $10^{12}$ cm$^{-2}$. However, it does not change the Raman G peak width or G peak frequency substantially when the Fermi level is within half the G phonon energy, corresponding to a charge density of $ca \pm 10^{12}$ cm$^{-2}$. Das *et al.* monitored dopants by Raman scattering using electrochemically top-gated graphene on SiO$_2$ and reported the correlation between G peak as well as 2D peak frequency and the charge density, showing only a weak 2D dependence below $10^{12}$ cm$^{-2}$.[18] Lee *et al.* analyzed the correlation of the 2D vs. G peak frequency and quantified the charge density by separating the strain line and the charge line in the correlation, and they also surmised that the charge vector requires charges $> 10^{12}$ cm$^{-2}$.[17] Therefore, the above methods are only suitable for evaluating graphene with a doping level higher than ~$10^{12}$ cm$^{-2}$.[19-22] For high purity samples, e.g., suspended graphene,[23] graphene encapsulated between hexagonal boron nitride (hBN) layers,[7] and graphene deposited on octadecyltrimethoxysilane (OTMS)[24], more precise optical methods still need to be developed to estimate the doping level of graphene. Here, we aim to develop an optical method that can evaluate the doping level and charge fluctuation in graphene in the range from $10^{10}$ cm$^{-2}$ to $10^{12}$ cm$^{-2}$ by using the split in the Raman 2D peak which appears at low doping levels.[25-27]

The 2D peak in the Raman spectra of graphene arises from the double resonance mechanism near the K-points between two nearby Dirac cones.[12, 28, 29] This resonant property makes it sensitive to any perturbation in the electronic states[12, 30] and phonon dispersion,[15, 28] thus the 2D mode can be used to monitor the doping level in graphene. Berciaud *et al.* investigated the asymmetric 2D line shape of suspended graphene with low charge density ($5 \times 10^{11}$ cm$^{-2}$) and found that the 2D peak can be separated into 2 peaks (2D$_1$ and 2D$_2$) in low doping regimes, with the 2D peak-split (difference between the 2D$_1$ peak frequency and the 2D$_2$ peak frequency) decreasing when the electro-static doping increases.[25, 26] For graphene sandwiched between hBN layers, the 2D peak-split at low accidental doping reveals a host of information on charge density screening of the electron and phonon dispersions in graphene.[31] However, a systematic study probing the 2D peak as a function of charge density is still lacking.

In this work, we use electrostatically gated graphene on OTMS-treated $SiO_2$/Si substrates to relate the split between the Raman $2D_1$ and $2D_2$ peaks with charge density and we compare the results to transport measurements. We found that the 2D peak-split indeed increases as the charge neutrality point (CNP) is approached, and that the absolute split at the CNP is dependent on the size of the local charge fluctuations. The 2D peak-split has a much larger variation with charge (~10 times) as compared to other Raman indicators and is therefore more sensitive to low charge variations. Our work provides a simple, noninvasive way to explore low doping levels and charge fluctuations in graphene with high precision, allowing evaluation of the graphene quality before fabricating graphene-based devices.

Clean graphene samples were prepared on OTMS self-assembled monolayers. The procedures of preparing substrates and treating OTMS on $SiO_2$/Si substrates are described in the Supplementary Information and Figure S1. Briefly, the $SiO_2$/Si substrates were cleaned and treated by OTMS using a spinning and vaporization method to achieve highly ordered self-assembled monolayer. The OTMS-treated $SiO_2$/Si substrate is smoother and more hydrophobic than a bare $SiO_2$ surface, with a root mean square (RMS) roughness of 0.188 nm, and a contact angle of 109° (Figure S2). The smooth and hydrophobic surfaces of the OTMS-treated substrates indicate the formation of a layer of highly-ordered OTMS molecules, which reduces dangling bonds and surface-adsorbed polar molecules.[32] Graphene samples were deposited on the OTMS-treated 90 nm $SiO_2$/Si substrate, identified by optical contrast under an optical microscope, and then confirmed by Raman spectroscopy.[30] Figure S3 shows the Raman G peak frequency and G peak width of graphene exfoliated on an OTMS-treated substrate versus graphene exfoliated on a bare SiO$_2$/Si substrate. OTMS effectively reduces the accidental doping level and helps maintain high-quality graphene, as gauged by the increase in the G peak width from 11.5 to 14 cm$^{-1}$.[16, 33, 34] Back-gated graphene devices were fabricated and the electrical properties of graphene was obtained by transport measurements. The cleanest graphene devices on OTMS-treated substrates had an accidental doping of $3.6 \times 10^{11}$ cm$^{-2}$, a charge fluctuation of $2.2 \times 10^{11}$ cm$^{-2}$, and a mobility of $\sim 1.7 \times 10^4$ cm$^2$/V·s, as deduced from fitting two-point probe transport measurements.

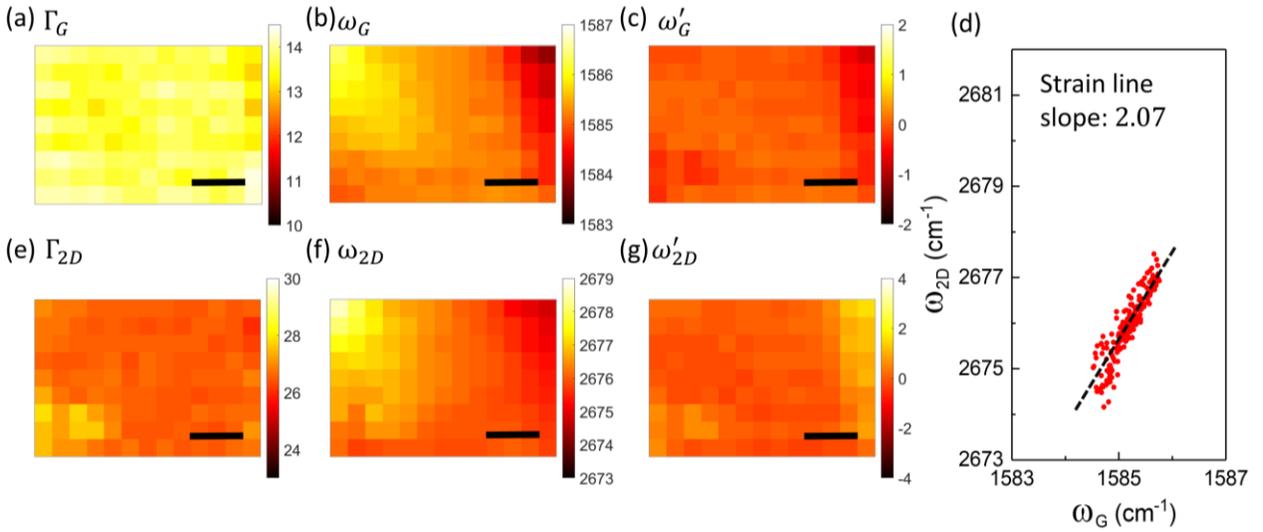

**Figure 1.** Raman spatial mapping of a representative graphene sample (9 μm by 13 μm) prepared on an OTMS-treated SiO$_2$/Si substrate. The scale bar is 3 $\mu m$. Raman map of the (a) G peak width ($\Gamma_G$), (b) G peak frequency ($\omega_G$), (c) G peak frequency after removing strain line ($\omega'_G$), (e) 2D peak width ($\Gamma_{2D}$), (f) 2D peak frequency ($\omega_{2D}$), (g) 2D peak frequency after removing strain line ($\omega'_{2D}$). (d) The G peak frequency versus the 2D peak frequency. The black dashed line is the fitted strain line. The $\omega_G$ vs. $\omega_{2D}$ data has a slope of 2.07.

Raman spatial mapping with a green laser (532 nm) was used to characterize the sample before the contacts had been laid down. Figure 1 shows the spatial Raman maps of a representative graphene sample on an OTMS-treated substrate and demonstrates homogeneity, apart from a correlated long-range "ridge" in the G and 2D peak frequencies (Figure 1b,f) which results from slight compression of the graphene sample.[9, 17, 35] A slight compressive strain is seen in Figure 1d, which shows the 2D versus G peak frequencies with a well-defined slope of 2.07, a typical value.[9, 17] As seen from Figure 1f, the variation in 2D peak frequencies is small enough that it is reasonable to ignore the strain-induced 2D peak-splitting.[36-39] We subtracted the strain line from the raw data to remove the effect of strain on the 2D and G frequencies, as shown in Figure 1c,g, and again, we found the maps to be homogeneous. The G peak width and G peak frequency has an average value of $\Gamma_G = 13.8 \pm 0.25$ cm$^{-1}$ and $\omega_G = 1585.3 \pm 0.4$ cm$^{-1}$, respectively. Figure S4 plots $\omega_G$ and $\Gamma_G$ versus doping levels and demonstrates that for doping below $0.5 \times 10^{12}$ cm$^{-2}$ there is no discernable variation with charge density.[40] Hence, the homogeneous distribution of the G peak width and G peak frequency under low doping condition only shows that the accidental doping is below this level. The Raman map of the 2D peak intensity over the G peak intensity ($I_{2D}/I_G$), as shown in Figure S5 is also homogeneous and give no information about lower doping levels. We will instead concentrate on the asymmetry of the 2D line, which is approximated by two symmetric peaks, 2D$_1$ and 2D$_2$.

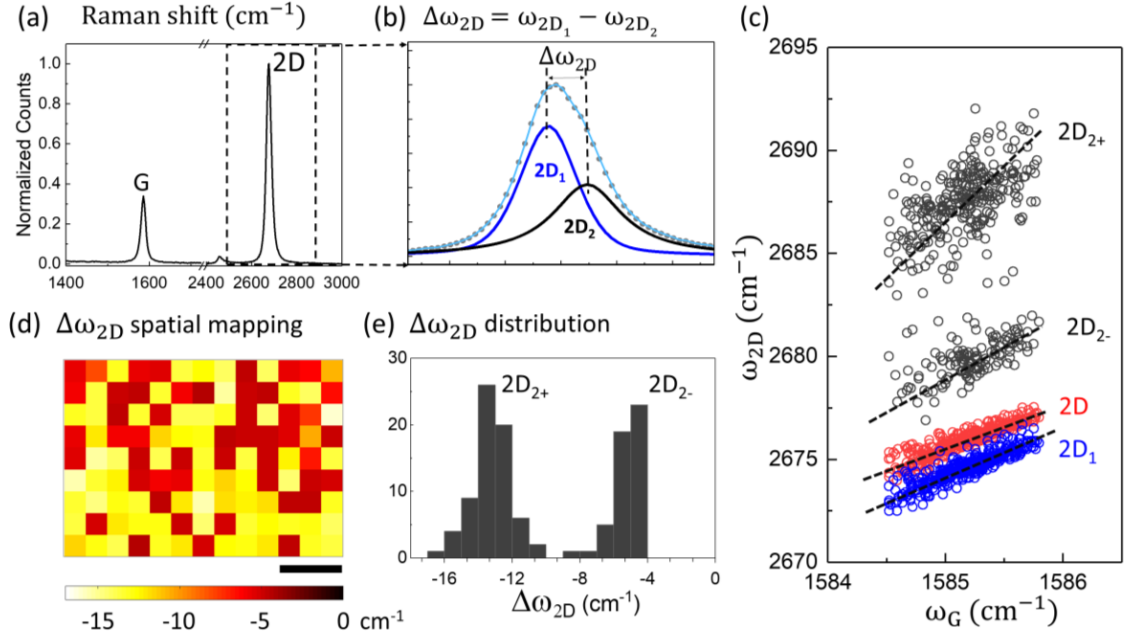

**Figure 2**. Raman 2D peak-split. (a) A representative Raman spectrum of the graphene on an OTMS-treated substrate. (b) The 2D peak (obtained from the dashed rectangle from (a)) is fitted by two Voigt profiles, $2D_1$ (blue curve) and $2D_2$ (black curve). The gray circles are the measured Raman data and the light blue curve is the sum of the $2D_1$ and $2D_2$ intensities. The 2D peak-split is denoted by $\Delta\omega_{2D} = \omega_{2D_1} - \omega_{2D_2}$. (c) The 2D (single fit, red circles), $2D_1$ (blue circles), and $2D_2$ (black circles) peaks are plotted versus the G peak frequency. The correlation shows the strain dependent slopes. (d) Raman spatial mapping and (e) a histogram of the 2D peak-split. The scale bar in (d) is 3 $\mu$m.

A representative Raman spectrum of a graphene sample on an OTMS-treated substrate is shown in Figure 2a. The asymmetric 2D peak at low doping level is fitted with two Voigt profiles with the lower energy peak denoted as $2D_1$ (blue curve) and the higher energy peak denoted as $2D_2$ (black curve), as shown in Figure 2b. We compared the results of fitting the 2D peak with one Voigt profile and two Voigt profiles, shown in Figure S6. The effect of fitting range on the fitting results is also shown in Figure S7 and Figure S8, which demonstrates the reliability and stability of our data processing process. The 2D peak-split is denoted by $\Delta\omega_{2D} = \omega_{2D_1} - \omega_{2D_2}$, which is a negative value. The origin of 2D$_1$ and 2D$_2$ is ascribed to the inner and outer process of the electron (hole) scattering from two nearby Dirac cones due to asymmetry in the electron and phonon dispersions.[31, 41] We plot the 2D$_1$ peak frequency, the 2D$_2$ peak frequency, and replot the 2D peak (single peak fit) versus the G peak frequency, shown in Figure 2c. The black dashed line is the linear fitting of the 2D (single fit) strain line with a slope of 2.07. The 2D$_1$ data is almost identical to the 2D single fit, albeit 1.5 cm$^{-1}$ lower, and has a larger strain slope of 2.24. Two distinct distributions in the 2D$_2$ vs. G peak frequency emerges: the higher energy band, 2D$_{2+}$, and the lower energy 2D$_{2-}$ which we assign to lower and higher doping levels, respectively (see discussion below). These two distinctly distributions can also be seen in Raman spatial maps, shown in Figure S9, which shows the Raman spatial mapping of peak width, peak area, peak frequency before and after removing strain factor of 2D$_1$ and 2D$_2$. The difference, $\Delta\omega_{2D}$ is plotted in Figure 2d-e and show large variations from 4-16 cm$^{-1}$. This should be compared to the maps of $\omega_G$, $\omega_{2D}$, $\Gamma_G$, $\Gamma_{2D}$ (Figure 1), and $I_{2D}/I_G$ (Figure S4) which show almost no variation. Hence, the 2D split has a much larger variation range (~ 10 times), indicating a higher sensitivity to charge density variation at doping levels below $10^{12}$ cm$^{-2}$. Figure S10 compares the variation of $\omega_G$, $\omega_{2D}$, $\Gamma_G$, $\Gamma_{2D}$, $I_{2D}/I_G$, and $\Delta\omega_{2D}$ from three different graphene samples prepared on OTMS treated substrates. All the results demonstrate that the 2D peak-split has a larger variation than other indicators.

Figure 3a shows the relation of the 2D peak-split versus the 2D$_1$ and the 2D$_2$ peak frequencies after removing the strain factor. Figure S11 compares the same correlation before and after removing the strain line. For both 2D$_1$ and 2D$_2$ peaks, there are two clusters above and below ~|8| cm$^{-1}$ 2D peak-split corresponding to lower and higher charge, respectively. These clusters have slightly different slopes, but for both high and low charge, the 2D$_1$ and 2D$_2$ slopes intersect at $\Delta\omega_{2D} = 0$ cm$^{-1}$ for all measured samples. From the samples that we analyzed, highly doped distributions always have larger values of intercepts at $\Delta\omega_{2D} = 0$ cm$^{-1}$.

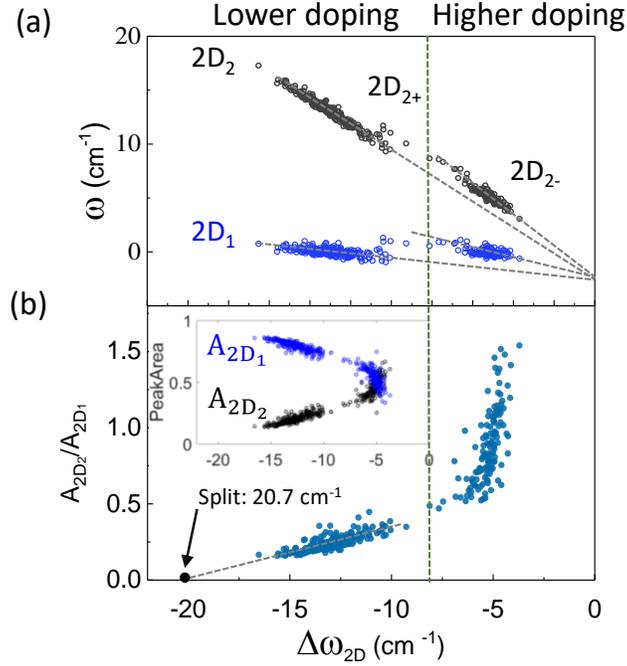

**Figure 3.** (a) The 2D$_1$ peak frequency (blue circles) and the 2D$_2$ peak frequency (black circles) after removing strain factor are plotted against the 2D peak-split. The gray dashed lines are linear fits of the different distributions. (b) The ratio of 2D$_2$ peak area over the 2D$_1$ peak area is plotted against 2D peak-split. The gray dashed line is the linear fitting of the data in the high split regime (low charge). The intercept of the fitted line corresponds to the 2D peak-split value of 20.7 cm$^{-1}$. The inset shows the plot of the normalized peak area of 2D$_1$ (blue circles) and 2D$_2$ (black circle) versus the 2D peak-split.

Figure 3b plots the ratio of the 2D$_2$ area over the 2D$_1$ area versus the 2D peak-split. This plot also reveals different slopes for lower and higher doping regimes. Concentrating on the low charge regime, the dashed line shows a linear fit with an intercept at $\Delta\omega_{2D} = 20.7$ cm$^{-1}$, which may correspond to the maximum split for zero charge in this sample. The inset in Figure 3 shows the variation of the normalized peak area of 2D$_1$ and 2D$_2$ versus the 2D peak-split, with 2D$_1$ dominating at high split regime (low doping level). There is a cross-over between the 2D$_1$ peak area and 2D$_2$ peak area when the split is at around 5 cm$^{-1}$. For a clean graphene sample such as suspended graphene and hBN-encapsulated graphene, a higher 2D$_1$ area has also been experimentally observed by previous work.[25, 31, 42]

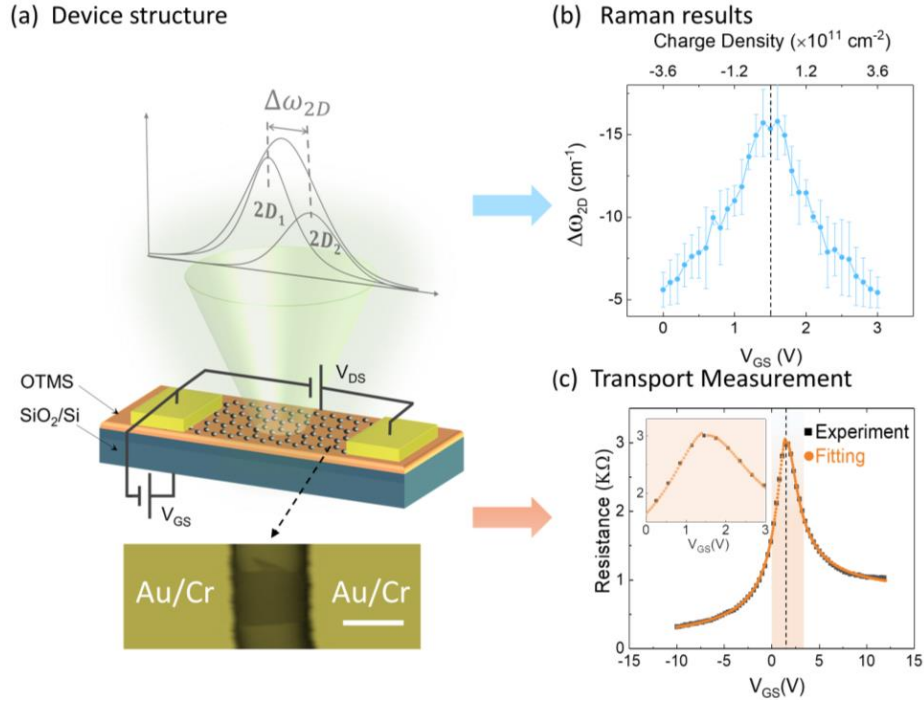

**Figure 4.** Monitoring graphene doping levels by the 2D peak-split. (a) Experimental setup of the GFET monitored by the Raman 2D peak-split. The top inset shows the measured Raman spectrum analyzed by the 2D peak-split ($\Delta\omega_{2D}$). The bottom inset shows an image of a representative GFET device, graphene connected between source and drain (Au/Cr) electrodes. Scale bar: 10 $\mu$m. (b) 2D peak-split as a function of back-gate voltage. The top axis shows the corresponding modulated charge density of the applied voltages. The charge neutrality point is at 1.55V, shown by dashed line in (c). The doping level of graphene was electrostatically tuned around the CNP to study the correlation between the split and the charge density at a lower doping level. (c) Transport measurement of GFET on an OTMS-treated substrate. The shaded region is the low doped regime (from 0 to 3V) that we focused on for the Gated Raman measurement, also shown in inset.

In order to quantify the relationship between the 2D peak-split ($\Delta\omega_{2D}$) and doping level ($n$), we electrostatically biased the graphene to induce a known charge density while measuring the 2D Raman response. In order to find the charge neutrality point (CNP), the charge density fluctuations ($\delta n$), and the mobility ($\mu$), we fabricated Graphene Field Effect Transistors (GFETs) to measure the transport curve.

Figure 4a shows the schematic of a GFET structure and the conceptual view of the GFET monitored by the Raman 2D peak-split. The back-gate voltage controls charge density and change the doping levels on graphene, making it possible to correlate Raman 2D peak-split as a function of charge density. For the GFET device fabrication, instead of using traditional photolithography which may contaminate the graphene sample, source and drain Au/Cr electrodes were deposited by e-beam evaporation using a shadow mask without touching the graphene and prevents contamination from photo-resist residues. The oxide surface was treated with OTMS to reduce charge puddling and increase mobility. Details about device fabrication and measurement are found in the Supplementary Information and Figure S1.

Figure 4c shows a plot of the channel resistance versus the back-gate voltage of the GFET device and the associated fitted curve. For the transport data shown in Figure 4c, the GFET has a channel aspect ratio of L/W = 1.5. The CNP due to the accidental hole doping is $V_g$ = 1.55V. The fitted $\mu$ was $\sim 1.7 \times 10^4$ cm$^2$/V·s, the fitted accidental doping ($n_0$) was $3.6 \times 10^{11}$ cm$^{-2}$, and the fitted $\delta n$ was $2.2 \times 10^{11}$ cm$^{-2}$. The doping level and the charge fluctuation are an order of magnitude lower than graphene samples deposited directly on SiO$_2$.[43, 44] This lower doping level makes it possible to study the Raman 2D response at different gating voltages below charge densities of $10^{12}$ cm$^{-2}$.

We used electrostatic back-gating of the graphene to study the Raman 2D peak-splits under different doping levels. Raman spectra were measured with various biasing values close to the CNP. Figure 4b shows the correlation between the 2D peak-split versus the back-gate voltage at a narrow doping range, $\pm 3.6 \times 10^{11}$ cm$^{-2}$ ($\pm 1.5$ V on the back gate). The result shows a strong correlation between the doping level (from 0 to $3.6 \times 10^{11}$ cm$^{-2}$) and the 2D peak-split (from 16 cm$^{-1}$ to 5 cm$^{-1}$). For both the hole and electron doping, the 2D peak-split reduces as the doping level of graphene increases. This strong correlation makes it possible to evaluate the charge density of graphene using the Raman 2D peak-split. By linearizing $\Delta\omega_{2D}$ versus $n$ dependent in the low doping regime ($|\Delta\omega_{2D}| > 8$ cm$^{-1}$), we find a variation of $2 \times 10^{10}$ cm$^{-2}$ per 2D peak-split wavenumber for this sample.

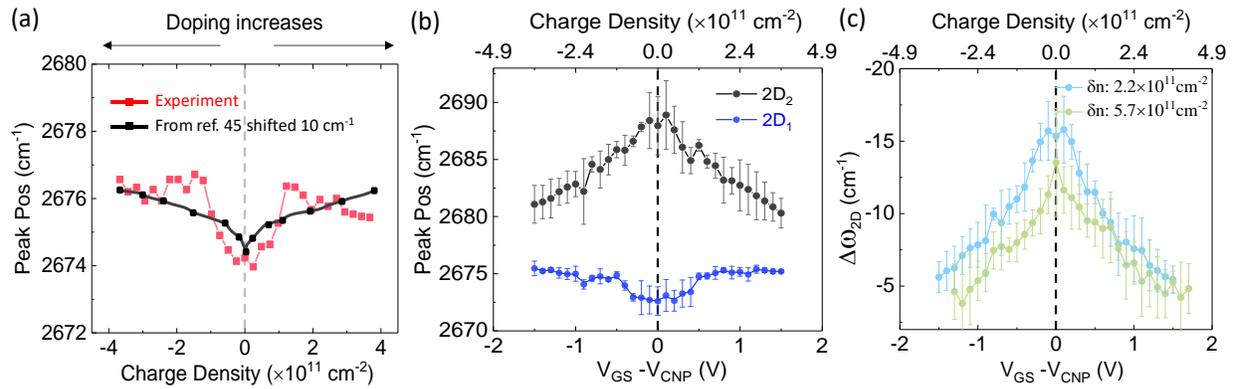

**Figure 5.** (a) The 2D peak frequency versus charge density. The red line is our experimental data and the black line is data extracted from ref.[45] (b) The relation of the 2D$_1$ peak (blue squares) frequency, and the 2D$_2$ peak (black squares) frequency versus the back-gate voltage. (c) The 2D peak-split as a function of the back-gate voltage. The blue line represents a graphene sample with a charge fluctuation of $\delta n = 2.2 \times 10^{11}$ cm$^{-2}$. The green line represents a graphene sample with a charge fluctuation of $\delta n = 5.7 \times 10^{11}$ cm$^{-2}$. The top axis in (b) and (c) show the corresponding modulated charge density due to the applied voltages.

We can next examine the role of charge fluctuations by looking at the correlation between the biased voltages and the 2D peaks, shown in Figure 5. Figure 5a shows the 2D (single fit) peak frequency versus the doping levels. For very clean suspended graphene, the 2D frequency has been shown to dip slightly near the CNP.[45] The plot shows a similar decrease in the 2D (single fit) peak near the CNP while not as sharp as the data from the suspended graphene. We ascribe the lack of sharpness to the local charge fluctuations in our devices.

Figure 5b shows the variation of the $2D_1$ and $2D_2$ peak frequencies at voltages around the CNP. The $2D_1$ and $2D_2$ peak frequencies shows very different dependencies on the back-gate voltage. The $2D_1$ frequency has minimal variation (< 2.5 cm$^{-1}$) under different doping levels and has its smallest value near the CNP, while the $2D_2$ position has prominent variation which strongly correlates to the doping level, a significant stiffening from 2680 cm$^{-1}$ to 2689 cm$^{-1}$ with its largest value around the CNP. Thus, the 2D peak-split mainly stems from the $2D_2$ peak shift as is also seen in Figure 2c and Figure 3a. Note that the decrease of $2D_2$ frequency has similar trend for both hole and electron doping. This makes the 2D peak-split a versatile tool to analyze the doping levels in graphene samples for both hole and electron doping.

Figure 5c shows the correlation of the 2D peak-split and the doping levels where the results from two devices with different charge density fluctuations are shown. Both plots show a strong correlation between the 2D peak-split and the back-gate voltage (doping levels). However, cleaner samples with smaller charge fluctuations ($2.2 \times 10^{11}$ cm$^{-2}$) have larger splits near the CNP than those seen in samples with larger charge fluctuations ($5.7 \times 10^{11}$ cm$^{-2}$). The transport measurement of the device with larger charge fluctuations is shown in Figure S12. We attribute the lower splitting to local charge puddles, even when the graphene is set to the CNP. While the charge fluctuation measured from transport measurements includes variation across the full channel, the Raman measurements are semi-local and are expected to have less charge fluctuations as compared to the entire channel. However, STM measurements have shown that the charge puddling size is on the nanoscale (~ 6-100 nm),[43, 44] which is much smaller than the spot size of the laser (~0.4 $\mu$m). Thus, a range of charge puddles were sampled within a Raman measurement.

The effect of the charge puddling reducing the apparent 2D peak-split can be understood by considering both the $2D_1$ and $2D_2$ peak frequencies and their intensity behavior as a function of charge. The $2D_1$ intensity dominates near CNP (Figure 3b) and the $2D_1$ peak itself moves much less than the $2D_2$ peak (Figure 5b). Hence, as a range of charge densities are sampled, the $2D_1$ will be strong and unmoving, while the $2D_2$ will be weakest for the maximum peak split and therefore higher charge densities will dominate the $2D_2$ signal and reduce the measured 2D peak-split. In our earlier work on graphene sandwiched between hBN,[31] with a charge fluctuation estimated at $4\times10^{10}$ cm$^{-2}$, we reported an estimated sensitivity of $1.1\times10^{10}$ cm$^{-2}$ per wavenumber peak split, which was based on typical values of residual charge and charge puddling in hBN-sandwiched graphene.

Given that the charge and dielectric screening affect both the electronic and phonon dispersion,[46] a different dielectric environment will change the details of the relationship between charge doping and the 2D peak-split. Furthermore, we note a qualitative difference between the "low" ($|\Delta\omega_{2D}| > 8$ cm$^{-1}$) and "high" ($|\Delta\omega_{2D}| < 8$ cm$^{-1}$) charge density regimes, as can be seen in Figure 3, indicating different charge screening behavior of the phonon and electron dispersions. Hence, while the 2D peak-split is a sensitive probe of the doping level, the absolute values need to be calibrated for graphene in different dielectric environments. The 2D peak-split is also laser dependent. For both graphene on OTMS/SiO$_2$ (this work) and for graphene sandwiched in hBN,[31] the 2D peak-split is larger for the 532 nm green laser line than the 633 nm red laser line.

For this work, we sought an optical probe that can be used to evaluate the low charge density and the low charge fluctuations across the graphene sample without the fabrication or processing steps necessary for transport measurements. Different variation of the parameters extracted from Raman spectra of graphene samples on OTMS-treated SiO$_2$/Si were analyzed, and the 2D peak-split was found to have much higher sensitivity at doping level below $10^{12}$ cm$^{-2}$ than other parameters such as $\omega_G$, $\omega_{2D}$, $\Gamma_G$, $\Gamma_{2D}$, and $I_{2D}/I_G$. Using a back-gated GFET, we measured the 2D peak-split versus doping level and found that the 2D peak-split can differentiate charge densities down to $2 \times 10^{10}$ cm$^{-2}$ per 2D peak-split wavenumber, two orders of magnitude higher precision than using G peak frequency and width[18], or the 2D versus G positions[17]. Our work provides a simple, noninvasive way to explore doping levels and charge fluctuation in graphene with high precision. The 2D peak-split method provides a platform for estimating the doping levels and the quality of a graphene sample before building a high-quality graphene device.

## ASSOCIATED CONTENT

**Supporting information**

Additional information on substrate preparation and the device fabrication process and measurements, comparison of Raman results between graphene on $SiO_2/Si$ and graphene on OTMS-treated $SiO_2/Si$ substrates, Ando model for calculating the G peak width and G peak frequency as a function of doping level, residual analysis of fitting 2D peaks with a single Voigt and two Voigt profiles, the effect of the fitting ranges on the 2D peak-split, comparison of variation of different parameters extracted from Raman spectra on other samples, Raman spatial mapping of $2D_1$ and $2D_2$, Raman results and transport measurements of GFET with larger charge fluctuations, and the fitting model for transport measurements of GFETs can be found in the Supplementary Information.

## AUTHOR INFORMATION

**Corresponding author**

*Email: swan@bu.edu

**Author Contributions**

A.K.S. directed the experimental measurements, Z.C. designed and performed the experiments, Z.C. and N.U. performed Raman data analysis, M.V. built the shadow mask and helped perform Raman measurements, Z.C. and D.B. developed the OTMS substrates coating procedures, A.K.S and Z.C. wrote the manuscript with input from all coauthors.

## ACKNOWLEDGMENT

We would like to thank Yuyu Li for fruitful discussion and Ang Liu for the help in device fabrication process. The authors gratefully acknowledge support from the United States National Science Foundation DMR grant 1411008.

# Monitoring the low doping regime in graphene using Raman 2D peak-splits: Comparison of gated Raman and transport measurements

*Zhuofa Chen†, Nathan Ullberg‡, Mounika Vutukuru†, David Barton† and Anna K Swan\*†‡+*

†Department of Electrical and Computer Engineering, Boston University, 8 St Mary's St, Boston Massachusetts 02215, United States of America.

‡Department of Physics, Boston University, 590 Commonwealth Ave, Boston Massachusetts 02215, United States of America.

+Photonics Center, Boston University, 8 St Mary's St, Boston Massachusetts 02215, United States of America.

S1. Substrate preparation and characterization.

Substrate gate preparation: The substrates were prepared by the following procedures to lay down gate contacts: A clean Si wafer with wet thermal $SiO_2$ (90 nm) was patterned by photolithography using photoresist S1813. The exposed $SiO_2$ was etched away by buffer oxide etchant (BOE) to be able to contact the Si layer. A back-gate gold pad was fabricated by e-beam evaporation using 5 nm Ti and 90 nm Au. The wafer was sonicated in acetone to remove the photo-resist and then diced into 8 mm by 8 mm chips.

Substrate surface treatment: The self-assembled monolayer (SAM) was deposited on the $SiO_2$ substrate by the spinning and vaporization method. The chip was firstly treated by oxygen plasma to enhance hydrophilicity. Subsequently, 10 μl of OTMS solution (obtained from Sigma– Aldrich) was pipetted on the chip and allowed to self-assemble for 20 s. The chip with OTMS solution was then spun at 3000 rpm for 10 s and put into a desiccator with ammonium hydroxide solution around the chip and left for 10 hours. The ammonium hydroxide solution facilitates the formation of the SAM. Finally, the substrates were sonicated in toluene for 5 min to remove residues. The schematic of substrate preparation process is shown in Figure S1a-f.

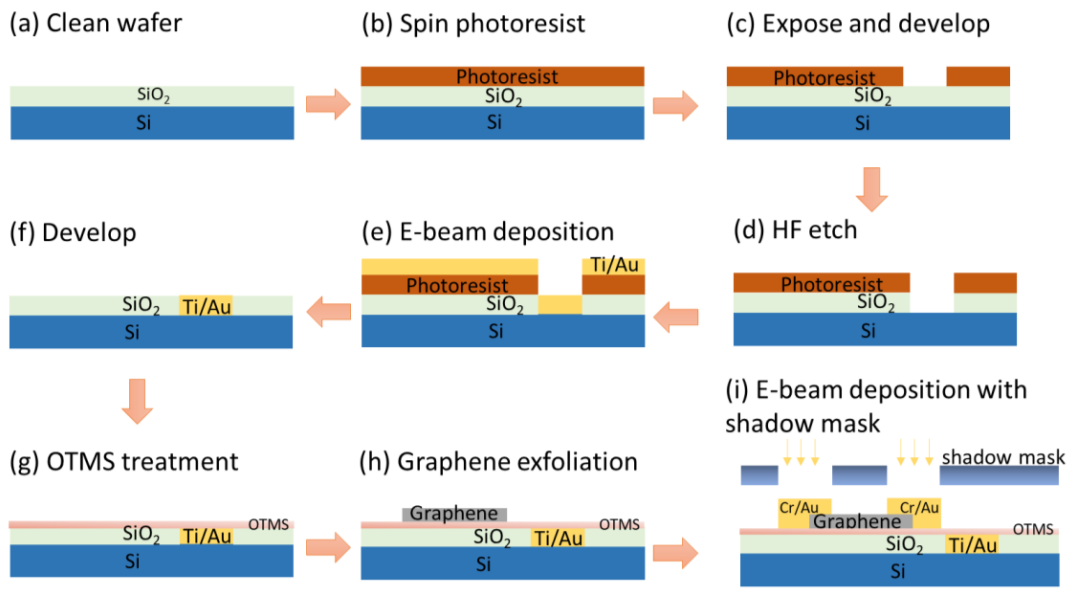

*Figure S1 Schematic of substrate preparation and device fabrication process.*

Figure S2a-b show the atomic force microscopy (AFM) surface topographic images of a representative $SiO_2$/Si substrate and a representative OTMS-treated $SiO_2$/Si substrate, respectively. The root mean square (RMS) is around 0.226 nm for $SiO_2$/Si substrates and 0.188 for OTMS-treated substrates. The scan region is 500 nm by 500 nm. The smooth surface of the OTMS-treated substrates is achieved by forming a layer of highly-ordered OTMS molecules, which reduces dangling bonds and surface-adsorbed polar molecules. The formation of the highly ordered OTMS surface can also be confirmed by the contact angle measurement, shown in Figure S2c-d. The advancing contact angle of the OTMS-treated surface is 109°, which is much larger than the contact angle (53°) of $SiO_2$/Si substrates. This highly hydrophobic property indicates the formation of compact hydrophobic groups of the OTMS molecules on the substrate.

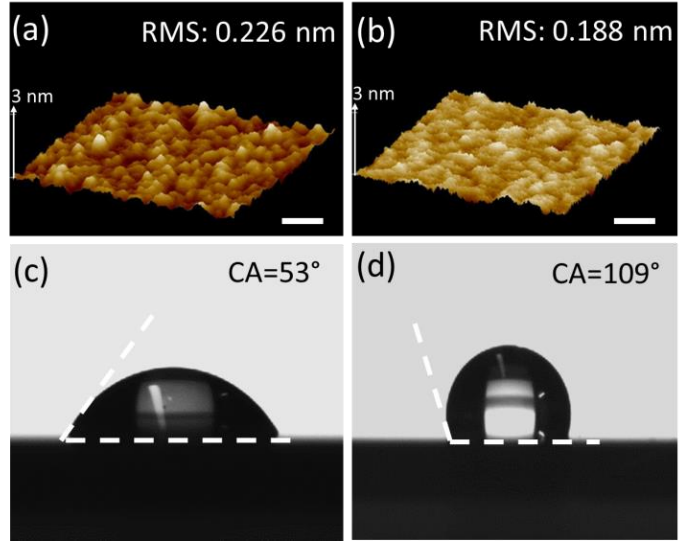

*Figure S2 Surface properties by Atomic Force Microscopy and contact angle measurements. The 3D AFM surface topographical images of (a) a SiO$_2$/Si substrate and (b) an OTMS-treated SiO$_2$/Si substrate. The RMS is 0.226 nm and 0.188 nm for the SiO$_2$/Si substrate and the OTMS-treated substrate, respectively. The scanning region is 500 nm by 500 nm. (c) and (d) The optical images of contact angle on a SiO$_2$/Si substrate and an OTMS-treated substrate. The measured contact angle is 53° and 109° for (c) the SiO$_2$/Si substrate and (d) the OTMS-treated substrate, respectively. The images reveal a smooth and highly hydrophobic surface of the OTMS-treated substrate. The scale bar in (a) and (b) is 100 nm.*

S2. Device fabrication and measurements.

Device fabrication: Graphene field-effect transistors (GFETs) on the OTMS-treated substrates were fabricated by a lithography-free process. Graphene samples were deposited on the OTMS-treated SiO$_2$/Si chip under ambient conditions using mechanical exfoliation and then identified by optical contrast and confirmed by Raman spectroscopy. Instead of using traditional photolithography which may introduce contamination on the graphene flakes, the source and drain Au/Cr electrodes were deposited by e-beam evaporation using a shadow mask, as shown in Figure

S1i. This was accomplished without touching the graphene sample. Cr was adopted to realize low contact resistance on the graphene samples.

Measurement: The Raman spectra were measured using the Renishaw Raman instrument with a green laser of 532 nm (2.33 eV). The beam size is 0.4 $\mu$m in diameter. The laser power is set to 2mW to prevent laser-induced thermal effects on samples during measurement. A 1200 grooves/mm grating was adopted for the spatial mapping of the Raman response. The collected optical radiation was dispersed onto the charged-coupled device (CCD) array with a spectral dispersion of 2 cm$^{-1}$ per pixel. The 2D peak, $2D_1$ peak, $2D_2$ peak, G peak were analyzed by a custom-built MATLAB (MathWorks) program. The 2D peak was fitted to a single Voigt profile (single fit for 2D peak) and two different Voigt profiles for $2D_1$ peak and $2D_2$ peak. The exposure time is adjusted to collect > 6k for 2D peak intensity. The IV characteristics of the GFETs were measured by a custom-built setup with two Keithley2400 power sources which were controlled by a MATLAB program for data acquisition and analysis. All of the transport measurements were carried out in a vacuum environment (~10 mTorr) at room temperature.

The graphene samples exfoliated on bare SiO$_2$/Si substrates and the graphene samples exfoliated on OTMS-treated substrates exhibit different properties in the Raman G peak frequency and G peak width, as shown in Figure S3. Compared to graphene on a bare SiO$_2$/Si substrate, graphene on an OTMS-treated surface exhibits a wider G peak width and a smaller G peak frequency, indicating lower more pristine graphene. The as-exfoliated graphene sample on an OTMS surface exhibited a larger G peak width (~14 cm$^{-1}$), which is comparable with previous reports on suspended graphene and BN/Graphene/BN stack. However, the graphene on a bare SiO$_2$/Si substrate generally sustains finite doping due to charged surface states and impurities, leading to a narrower (~11.5 cm$^{-1}$) line shape and a blue shift on the G peak due to compression (shown in Figure S3). We also found that graphene on OTMS shows a higher ratio of the peak intensity ($I_{2D}/I_G$)(data not shown here). It has been reported that the $I_{2D}/I_G$ would be reduced in doped graphene due to the electron-electron scattering interaction when the charge density increases.[40] The charged states and impurities increase the doping level of graphene on a $SiO_2$/Si substrate, which broadens the 2D peak and smears out the asymmetric property of the 2D peak. The results show that treating a $SiO_2$/Si substrate with OTMS can effectively reduce the accidental doping level and maintain the pristine quality of the graphene.

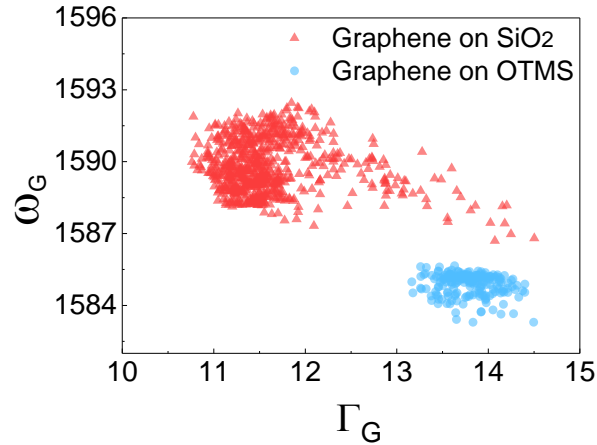

*Figure S3 Correlation between $\omega_G$ and $\Gamma_G$ for the graphene on an OTMS treated surface (blue dots) and graphene on bare $SiO_2$/Si substrate (red triangles). The spectra were measured by $E_L$=2.33 eV.*

S3. Ando-model: G peak width and G peak frequency versus doping levels.

The G peak frequency and G peak width can be described using the Ando model[47]:

$$\Pi(\varepsilon_0) = \lambda E_F - \frac{\lambda}{4}(\varepsilon_0 + i\delta)\left(\ln\left(\frac{\varepsilon_0 + 2E_F + i\delta}{\varepsilon_0 - 2E_F + i\delta}\right) + i\pi\right) \quad (S1)$$

Here $\lambda$ is the electron-phonon coupling and we use $\lambda = 8 \times 10^{-3}$, $E_F$ is the Fermi level of graphene, $\varepsilon_0 = 0.196$ eV (1582 cm$^{-1}$) is the G peak frequency of pristine graphene, which is unperturbed optical phonon energy of G peak, $\delta$ is the peak broadening factor describing electron level broadening. The G peak frequency and G peak width is given by:

$$\omega_G = \varepsilon_0 + Re\Pi(\varepsilon_0) \quad (S2)$$

$$\Gamma_G = \delta_0 + Im\Pi(\varepsilon_0) \quad (S3)$$

Figure S4 shows the calculated $\omega_G$ and $\Gamma_G$ as a function of doping levels. We can see that at low doping level around $\pm 4 \times 10^{11}$ cm$^{-2}$, shown as shaded region, the $\omega_G$ and $\Gamma_G$ almost remain constant as the Fermi level and charge density increase. This confirms our experimental results that show homogeneous Raman spatial mapping in Figure 1. We also see that graphene has a resonance at $E_F = \frac{\hbar\omega_G}{2}$, as shown in the Figure S4a, electrons are excited exactly to the Fermi level when an optical phonon decays. The shaded region is the low-doping regime that we concentrated on. Within this low doping regime, there is no discernable variation in G peak frequency and G peak width. While the 2D peak split shows much larger variation within this regime.

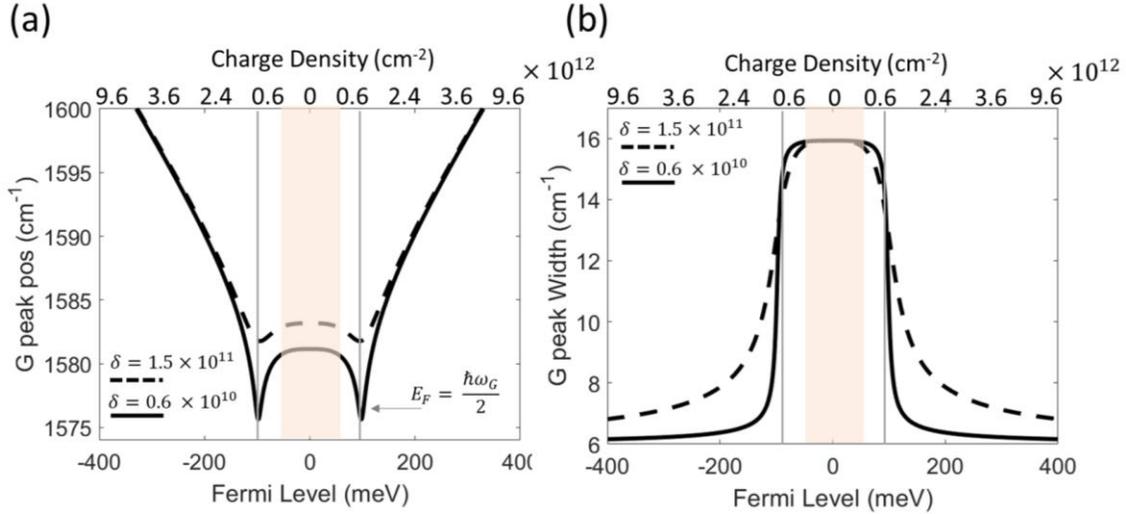

Figure S4 (a) G peak frequency as a function of fermi energy and charge density. The vertical gray line shows $E_F = \frac{\hbar\omega_G}{2}$, (b) G peak width as a function of fermi energy and charge density. The solid and dashed lines are calculated for $\delta = 0.6 \times 10^{11} cm^{-2}$ and $\delta = 1.5 \times 10^{11} cm^{-2}$, respectively. The shaded region is the low-doping regime that we are interested in the work, which shows no discernable variation in the G peak frequency and G peak width.

S4. Raman results and data analysis.

    S4.1. Raman mapping of 2D/G intensity.

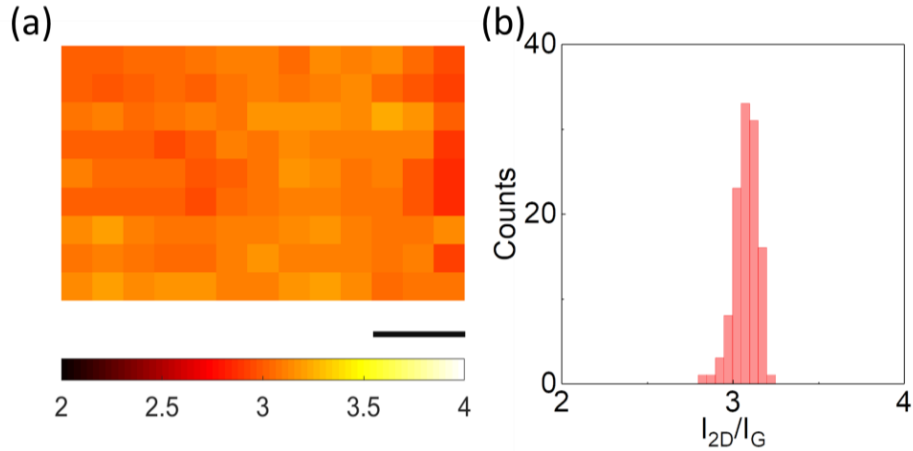

*Figure S5 (a) Raman spatial mapping of $I_{2D}/I_G$, scale bar: 3 μm. (b) A histogram of $I_{2D}/I_G$.*

Figure S5a shows a Raman map of the ratio of 2D peak / G peak intensity. We can see that the map shows spatial homogeneity. Figure S5b shows distribution of the $I_{2D}/I_G$ values, which mainly varies from 2.9 cm$^{-1}$ to 3.1 cm$^{-1}$ with a statistical spread of $3.07 \pm 0.07$ cm$^{-1}$. It has been reported that the ratio is larger in graphene with lower doping level and smaller in doped graphene due to the electron-electron scattering interaction when the charge density increases.[1,13,48]

S4.2.  Comparison of different fitting methods and fitting ranges for the Raman 2D peaks

Figure S6 compares the fitting results using one Voigt profile (one 2D peak) and two Voigt profiles (2D$_1$ and 2D$_2$) of a representative Raman spectrum taken from a graphene sample prepared on an OTMS-treated surface. Figure S6a shows the Raman data (gray circles) fitted using one Voigt profile (violet dashed line). It is clear that the 2D peak cannot fit the raw data very well due to the asymmetric profile of the data. The bottom plot shows the residuals from the fitting, which is calculated by subtracting the fitted data from the raw Raman data. The calculated residuals have an amplitude around 5% to 6% of the 2D peak intensity. A higher amplitude of the dashed curve indicates a larger residual, which corresponds to a worse fitting.  Figure S6b shows the same

Raman data fitted using two Voigt profiles $2D_1$ (blue line) and $2D_2$ (black line). The green dashed line is the sum of the $2D_1$ peak and the $2D_2$ peak. The results show that $2D_1$ and $2D_2$ can fit the Raman 2D peak well with negligible residuals as shown by the residuals plot on the bottom, which shows a much lower amplitude with a residual around 0.8% to 1% of the 2D peak intensity.

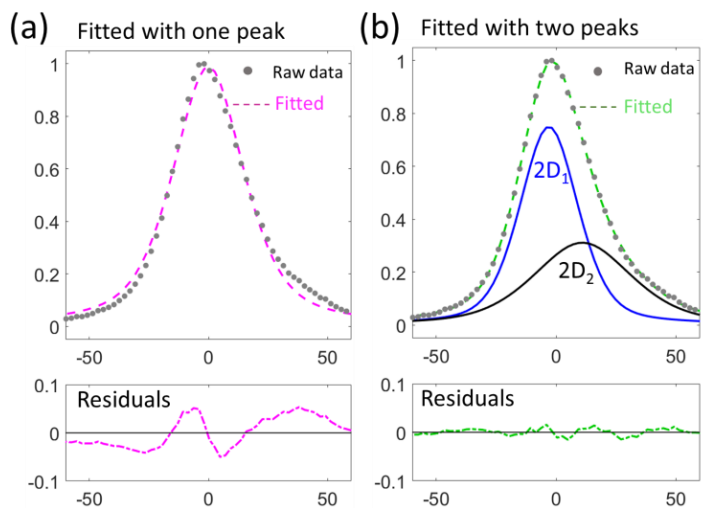

*Figure S6 Comparison of fitting the 2D peak with one Voigt profile and two Voigt profiles. (a) The 2D peak data (gray circle) is fitted using one Voigt profile (violet dashed line). The bottom plot shows the residuals calculated by subtracting the fitting data from the measured raw data. (b) The 2D peak data (gray circle) is fitted by two Voigt profiles ($2D_1$ is the blue line and $2D_2$ is the black line). The bottom plot shows the residuals of the fitting results. All of the amplitudes are normalized by the intensity of the 2D peak.*

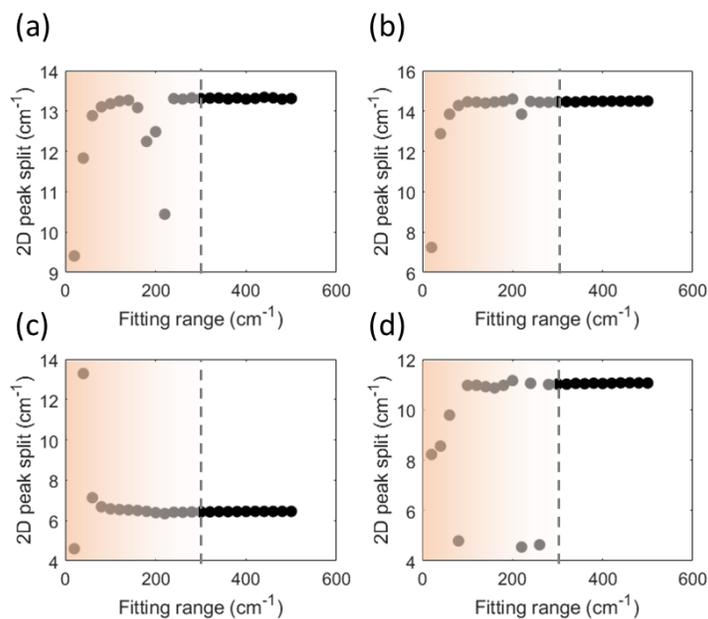

*Figure S7 The 2D peak split as a function of the fitting range. (a), (b), (c), and (d) correspond to four different graphene samples with a split of 13.3 cm$^{-1}$, 14.5 cm$^{-1}$, 6.5 cm$^{-1}$, 11.1 cm$^{-1}$, respectively. The unstable fitting regions are shaded on the left.*

We defined the range for fitting the 2D$_1$ and 2D$_2$ peaks as $\omega_0 \pm$ *fitting range*, where $\omega_0$ is the 2D peak frequency. We found that the fitting range indeed affects the fitting results (2D peak split), as shown in Figure S7. We tested different fitting ranges ranging from 20 cm$^{-1}$ to 500 cm$^{-1}$ on more than 10 different regions of different graphene samples. More than 200 spectra were analyzed. We found that the fitting result is not stable when the fitting range is less than 300 cm$^{-1}$ (as seen from the left of the dashed line in *Figure S7*a-d ) and the fitting results remain constant if the fitting ranges are larger than 300 cm$^{-1}$ (as seen from the right of the dashed line in *Figure S7*a-d). Therefore, all the data analysis in the paper using a fitting range of 300 cm$^{-1}$, making sure the accuracy and reliability of our analysis. *Figure S8* plots the fitting results of 2D$_1$ and 2D$_2$ peak area and peak width as a function fitting region, which also illustrates a fitting range of at least 300 cm$^{-1}$ should be guaranteed for processing the data.

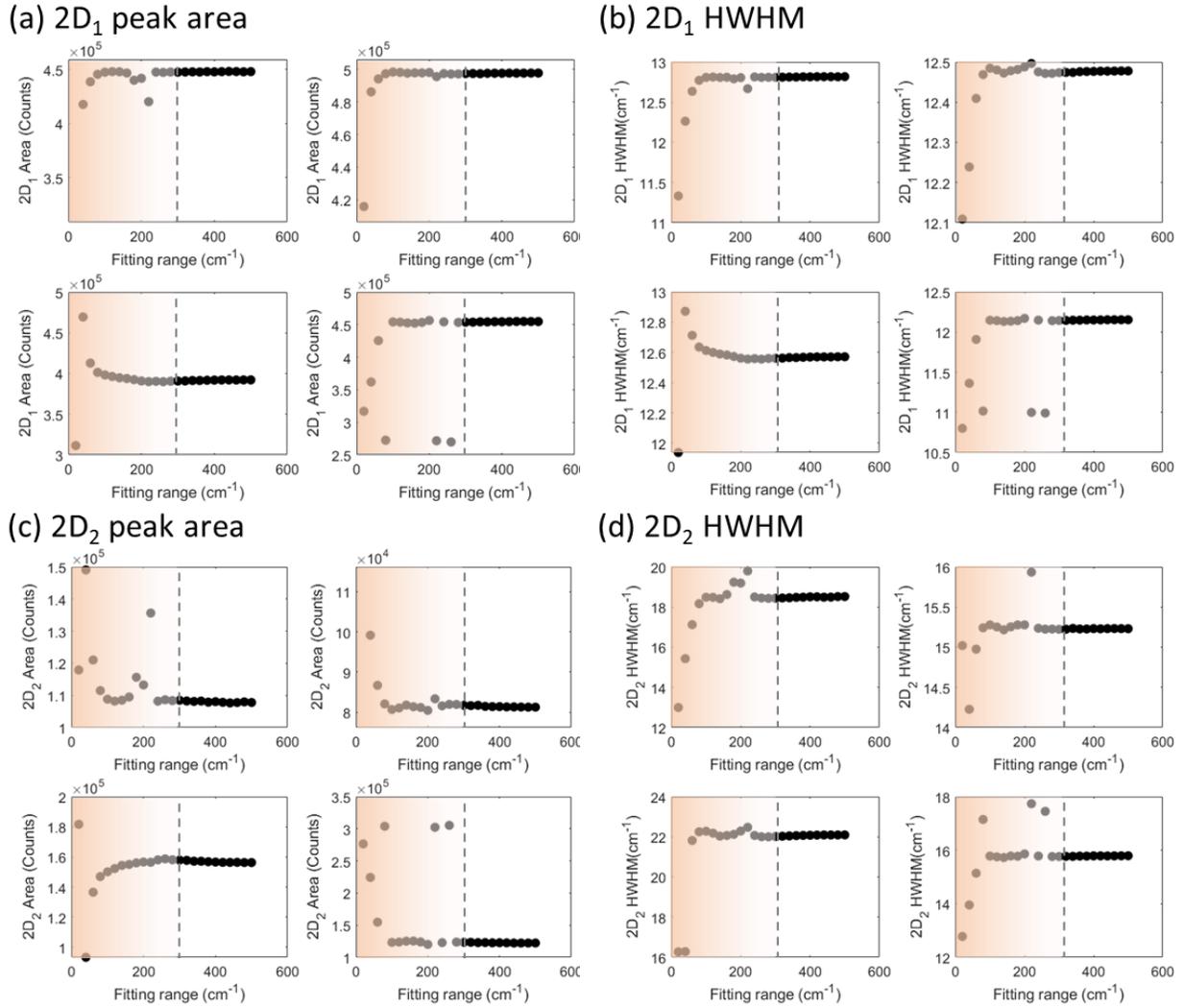

*Figure S8 Effect of the fitting range on the (a) 2D$_1$ peak area, (b) 2D$_1$ half width half maximum, (c) 2D$_2$ peak area, (d) 2D$_2$ half width half maximum in four different samples. The unstable fitting region are shaded on the left.*

S4.3. Raman spatial mapping of the 2D$_1$ peak and the 2D$_2$ peak

Figure S9 shows the Raman spatial map of the 2D$_1$ and 2D$_2$, including the peak width (FWHM), the peak area, and the peak frequency before and after removing the strain line of the 2D$_1$ and 2D$_2$, respectively. Figure S9a,e show the 2D$_1$ and 2D$_2$ peak widths ($\Gamma_{2D1}$ and $\Gamma_{2D2}$). We found that $\Gamma_{2D1}$ has less variation (mostly from 23 cm$^{-1}$ to 25 cm$^{-1}$) while $\Gamma_{2D2}$ has larger variation (spanning from 25 cm$^{-1}$ to 40 cm$^{-1}$). Figure S9b,f show spatial variations of the 2D$_1$ and 2D$_2$ peak area. As the 2D peak split increases, the peak area of 2D$_1$ increases and the peak area of 2D$_2$ decreases. The region with the maximum 2D$_1$ peak area corresponds to the region with the minimum 2D$_2$ peak area. Both the 2D$_1$ peak area and the 2D$_2$ peak area are normalized by their sum. Figure S9c,d,g,h show the 2D$_1$ and 2D$_2$ peak frequencies before and after removing the strain line. The strain line is fitted by the 2D versus G peak frequency. We can see that the 2D$_1$ has less variation within $\pm 1$ cm$^{-1}$ after removing the strain line. Meanwhile, the 2D$_2$ has a larger variation, 3 cm$^{-1}$ to 16 cm$^{-1}$ even after removing the strain line. The distance from the 2D$_1$ peak frequency and the 2D$_2$ peak frequency corresponds to the 2D peak-split. Thus, we can conclude that the variation of the split mainly due to the shift of the 2D$_2$ peak frequency.

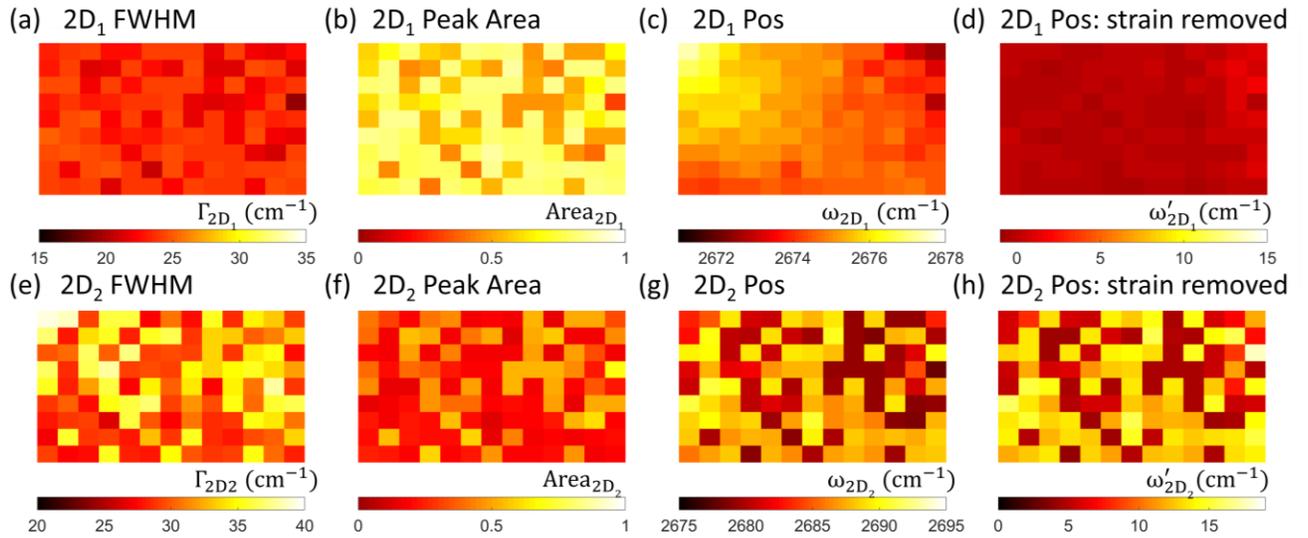

Figure S9 Raman spatial map of (a) the 2D$_1$ full width half maximum ($\Gamma_{2D_1}$), (b) the 2D$_1$ peak area ($Area_{2D_1}$, (c) the 2D$_1$ peak frequency ($\omega_{2D_1}$), (d) the 2D$_1$ peak frequency after removing the strain line ($\omega'_{2D_1}$), (e) the 2D$_2$ full width half maximum ($\Gamma_{2D_2}$), (f) the 2D$_2$ peak area ($Area_{2D_2}$), (g) the 2D$_2$ peak frequency ($\omega_{2D_2}$), (h the 2D$_2$ peak frequency after removing the strain line ($\omega'_{2D_2}$). The peak areas of 2D$_1$ and 2D$_2$ are normalized by their sum.

S4.4 Comparison of sensitivity due to charge density

Figure S10 compares the variation of $\omega_G$, $\Gamma_G$, $\omega_{2D}$, $\Gamma_{2D}$, $I_{2D}/I_G$, and the 2D peak split. Statistical data was extracted from Raman spatial mapping data of three different graphene samples prepared on OTMS-treated substrates. The variation is calculated as the standard deviation from all of the data points of each sample. We can see that the split has a much larger variation, 10 times larger than those of other parameters. This proves the sensitivity of using the 2D split to probe the charge density at low doping levels.

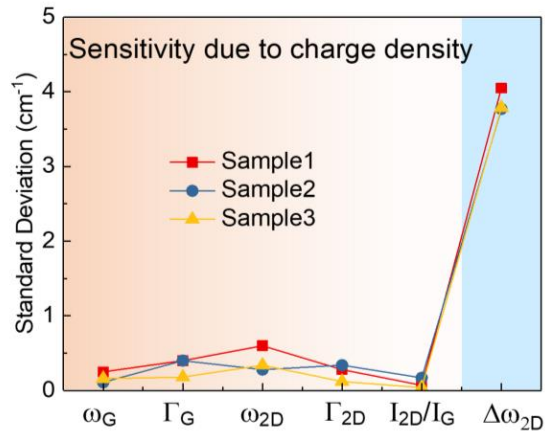

*Figure S10 Comparison of variation of different parameters extracted from Raman spectra, including the G peak frequency, G peak width, 2D peak frequency, 2D peak width, 2D intensity over G intensity, and the 2D peak split. The data is from three different graphene samples prepared on OTMS-treated substrates. The variation is defined as the standard deviation of each parameter.*

S4.5. Peak frequencies of $2D_1$ and $2D_2$ versus 2D peak-split

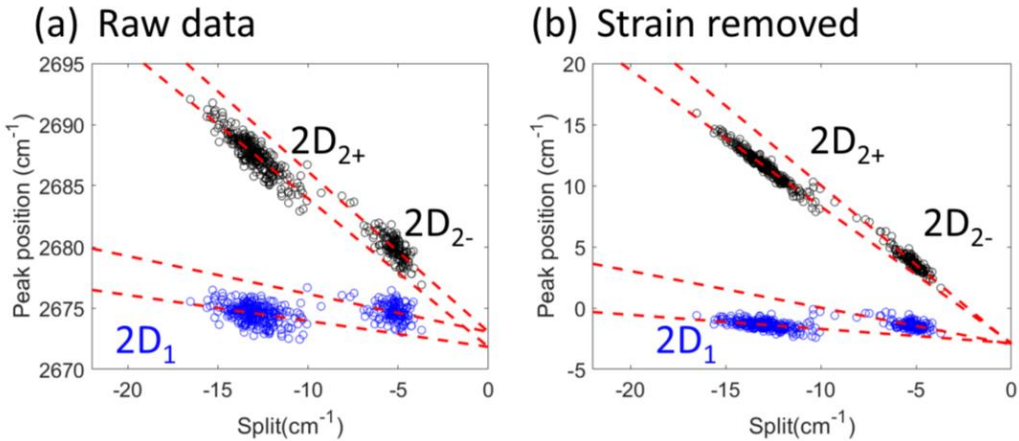

*Figure S11 Plot of the $2D_1$ peak frequency and the $2D_2$ peak frequency against the 2D peak split before (a) and after (b) removing the strain line. The four red lines are the linear fitted lines of the four distributions. The slopes and the intercepts of the fitted lines are shown in Table S1.*

Figure S11a shows the $2D_1$ peak frequency and $2D_2$ peak frequency as a function of the 2D peak split. The red lines are the corresponding linear fits of the data. We can see that the red lines of the highly doped and low doped distributions have intercepts at Split = 0. Figure S11b shows the same correlation before and after removing the strain factor, indicating the correlation is not affected by strain. The extracted parameters are summarized in Table S1. We did the same analysis on other graphene samples on OTMS-treated substrates and found similar results on all the other graphene samples (data not shown here).

Table S1 Fitted slopes and intercepts of the linear fitted lines of Figure S11

| Raw data | Fitted slope | Intercept (cm$^{-1}$) | Strain removed | Fitted slope | Intercept (cm-1) |
|---|---|---|---|---|---|
| 1 | -0.26 | 2671.14 | 1 | -0.11 | -1.38 |
| 2 | 1.26 | 2671.14 | 2 | -1.11 | -1.38 |
| 3 | -0.31 | 2673.07 | 3 | -0.29 | -1.59 |
| 4 | -1.31 | 2673.07 | 4 | -1.29 | -1.59 |

S5. Transport measurements of GFET with larger charge fluctuations.

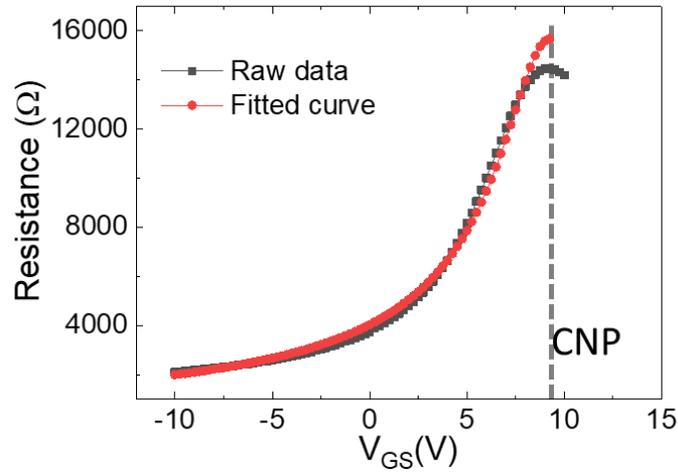

Figure S12 Transport curve of a GFET with a lower average 2D peak split. The black plot is the raw data and the red plot is the fitted curve. The fitted mobility is 5.2K $cm^2/V·s$, the accidental doping is $21 \times 10^{11} cm^{-2}$, and the charge fluctuation is $5.7 \times 10^{11} cm^{-2}$.

S6. Fitting model for transport measurements of GFET.

For a typical GFET device, the back-gate voltage can be described by equation S4.

$$V_G = \frac{E_F}{e} + \phi \qquad (S4)$$

Where $V_G$ is the back gate voltage, $E_F$ is the fermi level of graphene, e is the electron charge, and $e = 1.6 \times 10^{-19} C$. $\frac{E_F}{e}$ relates to the chemical capacitance of the graphene, and ϕ relates to the geometrical capacitance from the dielectric layer. $E_F$ and $\phi$ can be described by S5 and S6, respectively. [18]

$$E_F = \hbar |v_F| \sqrt{\pi n} \qquad (S5)$$

$$\phi = \frac{ne}{C_{OX}} \qquad (S6)$$

Where $\hbar$ is Planck's constant, $v_F$ is the fermi velocity, and n is the charge density on graphene due to gate biasing. $C_{OX}$ is the capacitance of the gate oxide dielectric layer. Note that though $E_F$ is highly dependent on temperature, we adopted a simplified model and get rid of the temperature factor in equation S2 as this has little effect on estimating the charge density introduced from back gate. The dielectric capacitance can be described using a simple parallel plate capacitor model, as shown in S7.

$$C_{OX} = \frac{\varepsilon_{SiO_2} \varepsilon_0}{d} \qquad (S7)$$

Where $\varepsilon_{SiO_2}$ is the dielectric constant of $SiO_2$, $\varepsilon_0$ is the vacuum permittivity, d is the thickness of the dielectric layer, and $C_{OX}$ is calculated to be $3.73 \times 10^{-8}$ F/cm² for the 90-nm dielectric layer in our device. We assume that the Fermi velocity is constant during the measurement and we adopted $v_F = 1.1 \times 10^6$ m/s. The relationship between the back-gate voltage and the charge density can thus be deduced:

$$V_G = \frac{\hbar |v_F| \sqrt{\pi n}}{e} + \frac{ne}{C} = 1.16 \times 10^{-7} \sqrt{n} + 4.1 \times 10^{-12} n. \qquad (S8)$$

For doping levels higher than $10^{10}$ cm$^{-2}$, the first part of S8 is negligible, thus

$$V_G \approx \frac{ne}{C}, \text{ or } n \approx 2.43 \times 10^{11} V_G. \qquad (S9)$$

We extracted the field effect mobility μ, the intrinsic doping $n_0$, and the charge fluctuation δn using the following equations: [49]

$$n_0 = C_{OX} V_{Dirac} \qquad (S10)$$

$$n_{total} = \sqrt{(\delta n)^2 + (C_{OX}|V_G - V_{Dirac}|)^2} \quad (S11)$$

$$R_{total} = R_C + \frac{L}{W}\frac{1}{\mu e n_{total}} \quad (S12)$$

$C_{OX}$ is the gate dielectric capacitance, $V_{Dirac}$ is the back gate voltage where graphene is at its CNP, and $n_0$ is the intrinsic doping level. $n_{total}$ is the total charge density, and $\delta n$ is the charge fluctuation due to residual impurities. The quantum capacitance is not included here because it is negligible compared to the carrier density modulated by the back gate. $R_{total}$ is the resistance between the source and drain electrodes, which includes the graphene channel resistance and the contact resistance $R_C$ (between graphene and the Au/Cr interface), L and W are the channel length and width, respectively, and μ is the electron or hole mobility in graphene.